\documentclass[runningheads,envcountsame]{llncs}
\usepackage{hyperref}
\usepackage{amsmath,amssymb}
\usepackage{xspace}
\usepackage{ifthen}
\usepackage{prooftree}
\usepackage{wrapfig}
\usepackage{listings}
\usepackage[all]{xy}

\hypersetup{%
  backref=true,  
  colorlinks=true,
  citecolor=blue,
  linkcolor=blue,
  anchorcolor=blue,
  filecolor=blue,
  menucolor=blue,
  urlcolor=blue
} 

\newcommand{\myvspace}[1]{\mbox{\hspace{0mm}}\vspace{#1}}

\newcommand{\semequality}{\;\mathtt{\equiv}\;}
\newcommand{\semeq}{\semequality}

\newcommand{\ttor}{$\mathtt{\mid}$}

{\marginpar{\textbf{End to see again !!!}}\par}

\ifx\undefined\note%
\newcommand{\note}[1]{\marginpar{{\tt #1}}}
\else
\renewcommand{\note}[1]{\marginpar{{\tt #1}}}
\fi

\newcommand{\A}{{\ensuremath{\mathbb A}}\xspace}
\newcommand{\B}{{\ensuremath{\mathbb B}}\xspace}

\newcommand{\SP}[1][none]{%
  \ifthenelse{\equal{#1}{none}}%
  {\Sigma\Pi}%
  {\Sigma\Pi(#1)}%
}

\newenvironment{mymatrix}{\left (\begin{matrix}}{\end{matrix}\right )}
\ifx\undefined\case
\newcommand{\case}{\medskip\noindent$\bullet\;\;$}
\else
\renewcommand{\case}{\medskip\noindent$\bullet\;\;$}
\fi

\newcommand{\<}{\langle}
\renewcommand{\>}{\rangle}

\newcommand{\etuple}{\<\>}

\newcommand{\ecotuple}{\{\}}

\newcommand{\rTo}[1][]{\xrightarrow{\;#1\;}}
\newcommand{\lTo}[1][]{\xleftarrow{\;#1\;}}

\newcommand{\Hom}[1]{\hom(#1)} 

\newcommand{\set}[1]{\{\,#1\,\}}

\newcommand{\code}[1]{\lstinline[basicstyle=\ttfamily]{#1}}
\newcommand{\cd}[1]{\code{#1}}

\title{%
  On the word problem for $\SP$-categories, \\
  and the properties of two-way communication\thanks{%
    Extended abstract. }
  \thanks{%
    Research partially supported
    by the ANR project SOAPDC no. JC05-57373.}
}%

\titlerunning{The word problem for $\SP$-categories}

\author{Robin Cockett\inst{1} and Luigi Santocanale\inst{2}}
\authorrunning{Cockett and Santocanale}

\institute{%
  Department of Computer Science, \\
  University of Calgary \\%
  \email{robin@ucalgary.ca}
  \and
  Laboratoire d'Informatique Fondamentale de Marseille, \\
  Universit\'e Aix-Marseille I, \\
  \email{luigi.santocanale@lif.univ-mrs.fr}
}
\date{\today}

\begin{document}
\maketitle


\begin{abstract}
  The word problem for categories with free products and coproducts (sums), 
  $\SP$-categories, is directly related to the problem of determining the 
  equivalence of certain processes. Indeed, the maps in these 
  categories may be directly interpreted as processes which communicate 
  by two-way channels.  

  The maps of an $\SP$-category may also be viewed as a proof 
  theory for a simple logic with a game theoretic intepretation.  The 
  cut-elimination procedure for this logic determines equality only 
  up to certain permuting conversions.  As the equality classes under 
  these permuting conversions are finite, it is easy to see that 
  equality between cut-free terms (even in the presence of the additive 
  units) is decidable.  Unfortunately, this does not yield a tractable 
  decision algorithm as these equivalence classes can contain exponentially 
  many terms.

  However, the rather special properties of these free categories -- and, 
  thus, of two-way communication --  allow one to devise a tractable algorithm 
  for equality.  We show that, restricted to cut-free terms  $s,t : X \rTo A$, 
  the decision procedure runs in time polynomial on $|X|\cdot|A|$, the 
  product of the sizes of the  domain and codomain type.

   \vskip 2pt

  {\bf Keywords.} $\SP$-categories, bicatersian categories, word
  problem, two-way communication, game semantics.
\end{abstract}


\section*{Introduction}


We present a decision procedure for equality of parallel arrows in
$\SP$-categories. These categories have (chosen) finite sums
(coproducts) and finite products, including, significantly, the units 
for these categorical operations.  Thus, the categories we consider do 
have an initial object, the unit for the sum, and a terminal object, 
the unit for the product.

Recall that word problems for algebraic theories amount to studying
the free models of these theories. Here the situation is analogous:
the theory of $\SP$-categories -- being an essentially algebraic
theory -- has free models; the decision procedure we describe 
relies crucially on a number of algebraic facts peculiar to free 
$\SP$-categories.

While the categorical structure we are investigating is one of the
simplest, the status of the word problem for these categories has
languished in an unsatisfactory state. It is decidable as standard 
tools from categorical logic \cite{lambek,dosen_book} allow free
$\SP$-categories to be viewed as deductive systems for logics. In 
\cite{CS00} these deductive systems were shown to correspond precisely 
to the usual categorical coherence requirements for products and sums 
and, furthermore, to satisfy the cut-elimination property.  The focus 
of the decision procedure then devolves upon the cut-free terms whose 
equivalence is completely determined by a finite number of ``permuting 
conversions''.

The cut-free terms, which represent arrows between two given types, are 
finite in number and this implies, immediately, that equality is decidable.
However, the implied complexity of this way of deciding equality is 
exponential because there can be an exponential number of equivalent terms. 
The question, which still remained open, was whether the matter could be 
decided in polynomial time. This was of particular interest as these 
expressions are, in the process world, the analogue of Boolean expressions.   
The main contribution of this paper is to confirm that there is a polynomial 
algorithm which settles this question.


There have been, directly or indirectly, a number of contributions towards our
goal in this paper.  Most of them involve a representation theorem, that is
the provision of a full and faithful functor from some variant of the free 
$\SP$-category into a concrete combinatoric category. For example
 \cite{joyal_hu} considers $\SP$-categories, in which the initial and 
final object coincide and represents these using a subcategory of the 
category of coherent spaces, while \cite{glabbeek} and \cite{dosen_last} 
both gives a representation of $\SP$-categories {\em without units} into, 
respectively, a combinatoric category of proof-nets and the category of sets 
and relations.  These related results, however, work only for the 
fragment without units -- or, more precisely, for the fragment with a common 
initial and final object.  As far we know, there is no representation theorem 
for the full fragment with distinct units. 

Units add to the decision problem -- and to the representation theory
-- a non-trivial challenge which is easy to under estimate. In particular, in \cite{CS00}, 
one of the current authors was guilty of rather innocently proposing an altogether too 
simple decision procedure which, while working perfectly in the absence of units, fails 
manifestly in the presence of units.  The effect of the presence of units on the setting 
is quite dramatic. In particular, when there are no units (or 
there is a zero) {\em all\/} coproduct injections are monic. However, rather contrarily, in 
the presence of distinct units this simply is not longer the case. Furthermore, this can be 
demonstrated quite simply, consider the following diagram:
$$
\xymatrix{0\times 0  \ar@<0.2em>[rr]^{\pi_{1}} \ar@<-0.2em>[rr]_{\pi_{0}} && 0
  \ar[rr]^{\sigma_{0}} && 0 + 1}
$$
As $0 + 1 \simeq 1$ is a terminal object, there is at most one
arrow to it: this makes the above diagram a coequalizer. Yet, the arrows
$\pi_{0},\pi_{1}$ are distinct in a free $\SP$-category, as for
example they receive distinct interpretations in the dual category of
the category of sets and functions.

As logicians and category theorists, we were deeply frustrated by this failure to 
master the units. The solution we now present for this decision problem, however, 
was devised only after a much deeper algebraic understanding of the structure of free
$\SP$-categories had been obtained. The technical observations which underly this development, 
we believe, should be of interest to logician and category theorists alike.
Yet, our principal motivation for studying the theory of
$\SP$-categories and free $\SP$-categories arose from the role they
have as models of computation. We discuss in details this point next.


The proof theory of (free) sums and products in a category is remarkable 
from a number of points of view.  Not only does it provide an elegant proof
theory with a rich underlying algebraic calculus, but also it supports
a variety of quite surprising interpretations.

The most immediate interpretation, but by no means one which is
transparently obvious see \cite{blass,joyal_money_games,luigis}, is as
a game theory in which the types represent finite games.  The products
have the role of opponent while the sums have the role of player and
there is no requirement that plays alternate.  The maps are then
interpreted as being mediators between games which use the information
of one game to determine the play on the other.  Their composition is
given by hiding the transfer of information which happens through
moves on a middle game.

Proof theoretically this composition can be viewed as a
cut-elimination process which, in turn, algebraically translates into
an elegant reduction system which is confluent modulo equations (the
details are to be found in \cite{CS00}).  As shall become clear this
paper is largely concerned with the consequences of the equations
which remain after the cut has been eliminated.  However, before
discussing this we must describe a second important and appealing
interpretation.

Arrows in the free category with sums and products can also be
interpreted as processes which communicate along channels: the types
are the (finite) protocols which govern the interactions along these
channels. These protocols tell a process which wishes to communicate
along a channel whether it is the turn of the
process to send a message (and precisely which messages can then be
sent) or whether it is the turn of the process to listen (and
precisely which messages can be received). This is more than an idle 
idea: the theoretical details of this interpretation have been fleshed 
out in some detail (and, in fact, more generally to allow multiple 
channels) in \cite{pastro2,pastro3}.

This last interpretation is quite compelling as the algebraic results
described in this paper suggest a number of not very obvious and even
somewhat surprising properties of communication along a channel. For
example, a process which is required to send a value could send
various different values and yet, semantically, remain {\em exactly
  the same process}. This is the notion of \emph{indefiniteness} which
is central in the business of unraveling the meaning of
communication. There are various situations in which this apparently
unintuitive situation can arise.  For example, it could be that the
recipient of the communication has simply stopped listening.  It is of
course very annoying when this happens but, undeniable, this is an
occurrence well within the scope of the human experience of
communication.  However, it can also be, more dramatically, that the sender has 
stopped communicating to the receiver -- and this produces what we shall call a 
\emph{disconnect}.

Proof theoretically and algebraically this all has to do with the
behavior of the (additive) unit, that is, the final object and the
initial object.  The purpose of this paper is to focus on these units
and their ramifications in the whole business of communication.  It is
certainly true that without the units the situation is very much
simpler.  However, if one is tempted therefore simply to omit them, it
is worth realizing that without any units there is simply no
satisfactory notion of {\em a finite communication}!

Of course, without the units the theory is not only simpler but a good
deal less mathematically interesting.  It is this mathematics which we
now turn to.

The paper is structured as
follows. In Section \ref{sec:deductivesystem}, we recall the
elementary definition of $\SP$-category, and the results of
\cite{CS00}. In Section \ref{sec:softness}, we start our analysis of
the main property of $\SP$-categories, softness, in order to give it a
more concrete meaning, accessible to the general logician, in terms of
a sort of undirected rewrite system. In Section
\ref{sec:weakdisjointness}, we shall present our first main result,
stating that coproduct injections are weakly disjoint in free
$\SP$-categories, and list some consequences. This leads to a discussion 
of arrows which factor through a unit -- indefinite arrows -- which
play a key role in the decision procedure. In Section
\ref{sec:bouncing} we present our second main observation: if two 
arrows in $\hom(X \times Y,A)$ and $\hom(Y,A+B)$ are definite but are 
made equal when, respectively, projecting and coprojecting into 
$\hom(X \times Y,A + B)$, this fact is witnessed by a unique ``bouncer'' 
in $\hom(Y,A)$. In Section \ref{sec:dp}, we collect our observations and 
sketch the decision procedure.



\section{The construction of free $\SP$-categories}
\label{sec:deductivesystem}

\subsection{$\SP$-categories}

We invite the reader to consult \cite{maclane}
for the basic categorical notions used in this paper. Here, an
\emph{$\SP$-category} shall mean a category with finite products
and finite coproducts.

Recall that a category has \emph{binary products} if, given two
objects $A,B$, there exists a third object $A \times B$, and natural
transformations
\begin{align*}
  \Hom{X,A}\times \Hom{X,B} & \rTo[\langle \;,\;\rangle] 
  \Hom{X,A\times B}\,,
  \\
  \Hom{X_{i},A} & \rTo[\pi_{i}]  \Hom{X_{0}\times X_{1},A} 
  \,, \;i = 0,1\,,
\end{align*}
that induce inverse bijections:
\begin{align*}
  \pi_{i}(\langle f_{0},f_{1}\rangle) 
  & = f_{i}\,,\;i = 0,1\,,
  & \langle \pi_{0}(f),\pi_{1}(f)\rangle & = f\,. 
\end{align*}
A \emph{terminal object} or \emph{empty product} in a category is an
object $1$ such that, for each object $X$, $\Hom{X,1}$ is a
singleton. It is part of standard theory that a terminal object is
unique up to isomorphism and that it is the unit for then product, as
$X \times 1$ is canonically isomorphic to $X$.

We obtain the definition of binary sums (or coproducts) and of initial
object, by exchanging the roles of left and right objects in the
definition of products: a category has \emph{binary sums} if, given
two objects $X,Y$, there exists a third object $X + Y$ and natural
transformations
\begin{align*}
  \Hom{X,A}\times \Hom{Y,A} & \rTo[\{ \;,\;\}] 
  \Hom{X + Y,A}\,,
  \\
  \Hom{X,A_{j}} & \rTo[\sigma_{j}]  \Hom{X,A_{0} + A_{1}} 
  \,, \;j = 0,1\,,
\end{align*}
that induce inverse bijections:
\begin{align*}
  \sigma_{j}(\{ f_{0},f_{1}\}) 
  & = f_{j}\,,\;j = 0,1\,,
  & \{ \sigma_{0}(f),\sigma_{1}(f)\} & = f\,. 
\end{align*}
An \emph{initial object} $0$ is such that, for each object $A$,
$\Hom{0,A}$ is a singleton.

\vspace{-12pt}
\begin{wrapfigure}[8]{r}{0pt}
  $$
    \xymatrix{
      \A \ar[rrdd]^{F}\ar[rr]^{\eta}& & \SP[\A]\ar@{{}{.}{>}}[dd]^{\exists!
        \,\tilde{F}} \\
      \\
      && \B
    }
  $$
\end{wrapfigure}
\myvspace{12pt}\par
A functor between two
$\SP$-categories $\A,\B$ is a \emph{$\SP$-functor} if it sends
(chosen) products to products, and (chosen) coproducts to
coproducts. 
The \emph{free $\SP$-category over a category $\A$}, denoted $\SP[\A]$,
has the following property: there is a functor $\eta : \A \rTo
\SP(\A)$ such that, if $F: \A \rTo \B$ is a functor that
``interprets'' $\A$ into a $\SP$-category $\B$, then there exists a
unique $\SP$-functor $\tilde{F} : \SP(\A) \rTo \B$ such that
$\tilde{F}\circ \eta = F$. This is the usual universal property
illustrated by the diagram on the right.

\smallskip

The free $\SP$-category on $\A$, can be ``constructed'' as follows.
Its objects are the types inductively defined by the grammar
\begin{align}
  \label{grammar:types}
  T & = \eta(x) \mid 1 \mid T \times T \mid 0 \mid T + T\,,   
\end{align}
where $x$ is an object of $\A$.  Then proof-terms are generated
according to the deduction system of figure
\ref{fig:deductive_system}. Finally, proof-terms $t: X \rTo A$ are
quotiented by means of the least equivalence relation that forces the
equivalence classes to satisfy the axioms of a $\SP$-category.
\begin{figure}[h]
  \centering
  \small
  $$
  \begin{array}[t]{|@{\hspace{5mm}}c@{\hspace{10mm}}c@{\hspace{5mm}}|}
    \hline
    &\\
    \begin{prooftree}
      \using{\text{identity-rule}}
      \justifies
      X \rTo[id_{X}] X 
    \end{prooftree}
    &
    \begin{prooftree}
      X \rTo[f] C
      \;\;\;\;C \rTo[g] A
      \using{\text{cut-rule}}
      \justifies
      X \rTo[f;g] A
    \end{prooftree} 
    \\
    &\\
    \hline
    &\\
    \multicolumn{2}{|c|}{%
      \begin{prooftree}
        x \rTo[f] y
        \using{\text{Generators rule}}
        \justifies
        \eta(x) \rTo[\eta(f)] \eta(y)
      \end{prooftree} 
    }%
    \\
    &\\
    \hline
    &\\
    
    &
    \begin{prooftree}
      -
      \using{R1}
      \justifies
      X \rTo[!] 1
    \end{prooftree}\\[12pt]
    
    \begin{prooftree}
      X_{i} \rTo[f] A
      \using{L_{i}\times}
      \justifies
      X_{0} \times X_{1} \rTo[\pi_{i}(f)] A 
    \end{prooftree}
    &
    \begin{prooftree}
      X \rTo[f] A \;\;\;\; X \rTo[g] B
      \using{R\times}
      \justifies
      X \rTo[\langle f,g\rangle] A \times  B
    \end{prooftree} \\
    &\\
    \hline
    &\\
    
    \begin{prooftree}
      -
      \using{L0}
      \justifies
      0 \rTo[?] A
    \end{prooftree} & \\[12pt]

    \begin{prooftree}
      X \rTo[f] A \;\;\;\; Y \rTo[g] A
      \using{L+}
      \justifies
      X + Y \rTo[\{ f,g\}] A
    \end{prooftree}
    &
    \begin{prooftree}
      X \rTo[f] A_{j}
      \using{R_{j}+}
      \justifies
      X \rTo[\sigma_{j}(f)] A_{0} + A_{1} 
    \end{prooftree}
    \\
    &\\
    \hline
  \end{array}
  $$
  \caption{The deductive system for $\SP[\A]$}
  \label{fig:deductive_system}
\end{figure}
Of course, while this is a perfectly good specification, we are
looking for an effective presentation for $\SP[\A]$. A first step in
this direction comes from the fact the identity-rule as well as the
cut-rule can be eliminated from the system.  More precisely we have
the following theorem:
\begin{proposition}[See \cite{CS00} Proposition 2.9]
  The cut-elimination procedure gives rise to a rewrite system that is
  confluent modulo the set of equations of figure \ref{fig:equations}.
\end{proposition}
\begin{figure}[h]
  \centering
  $$
  \begin{array}{|@{\hspace{5mm}}rcl@{\hspace{5mm}}rcl@{\hspace{5mm}}|}
    \hline
    \multicolumn{6}{|c|}{} \\
    \pi_{i}(\langle f,g\rangle)
    & = & \langle \pi_{i}(f),\pi_{i}(g)\rangle 
    &
    \sigma_{j}(\{ f ,g \}) & = & 
    \{ \sigma_{j}(f) ,\sigma_{j}(g) \} \\[12pt]
    \multicolumn{6}{|c|}{%
      \pi_{i}(\sigma_{j}(f))  = \sigma_{j}(\pi_{i}(f))
    }\\[12pt]
    \multicolumn{6}{|c|}{%
      \{\langle f_{11},f_{12}\rangle,
      \langle f_{21},f_{22}\rangle
      \}  =  \langle \{f_{11},f_{21}\},\{f_{12},f_{22}\}\rangle%
    }
    \\[12pt]
    \pi_{i}(!) & = & !
    &
    \sigma_{j}(?) & = & ? \\[12pt]
    \{ !,! \} & = & ! &
    \langle ?,? \rangle & = & ? \\[12pt]
    \multicolumn{6}{|c|}{!_{0}  =  ?_{1}}\\
    \multicolumn{6}{|c|}{} \\
    \hline
  \end{array}
  $$
  \caption{The equations on (identity\ttor cut)-free proof-terms}
  \label{fig:equations}
\end{figure}
From this we obtain an effective description of the category
$\SP[\A]$: the objects are the types generated by the
grammar \eqref{grammar:types}, while the arrows are equivalence classes
of (identity\ttor cut)-free proof-terms under the least equivalence
generated by the equations of figure \ref{fig:equations}. Composition
is given by the cut-elimination procedure, which by the above theorem
is well defined on equivalence classes.

Thus, our main goal in the rest of the paper is the following:
\textsf{given two proof-terms $s,t : X \rTo A$, are they equivalent
  according to the least equivalence relation generated by the
  equations of figure \ref{fig:equations}}? The problem
is easily seen to be decidable: the contribution of this paper is to 
show that, furthermore, there is a feasible algorithm. 

The main theoretical tool we shall use in developing this algorithm is 
the idea of {\em softness} which we now introduce.  In every $\SP$-category
there exist canonical maps 
\begin{equation}
  \label{diag:atomicity}
  \begin{split}
    \coprod_{j} \Hom{X,A_{j}}  \rTo[\;\;\;\;] & \Hom{X,\coprod_{j}
      A_{j}} \,,\\ 
    \coprod_{i} \Hom{X_{i},A} & \rTo[\;\;\;\;] \Hom{\prod_{i}
      X_{i},A}\,.
  \end{split}
\end{equation}
We shall be interested in these maps when, in a free 
$\SP$-category $\SP[\A]$, $X = \eta(x)$ and $A = \eta(a)$ are generators.
\myvspace{36pt}\\
In every $\SP$-category
there also exist canonical commuting diagrams of the form
\begin{align}
  \label{diag:softness}
  & \xygraph{%
    []*+{\coprod_{i,j} \Hom{X_{i},A_{j}}}
    (
    :[rrrr]*+{\coprod_{j} \Hom{\prod_{i} X_{i},A_{j}}}
    :[d(1.5)]*+{\Hom{\prod_{i} X_{i},\coprod_{j}A_{j}}}="E"
    )
    :[d(1.5)]*+{\coprod_{i} \Hom{ X_{i},\coprod_{j}A_{j}}}
    :"E"
  }
\end{align}
The following key theorem holds:

\begin{theorem}[See \cite{CS00} Theorem 4.8]
  \label{theo:characterization}
  The following properties hold of $\SP[\A]$:
  \begin{enumerate}
  \item The functor $\eta : \A \rTo \SP[\A]$ is full and
    faithful.
  \item Generators are \emph{atomic}, that is, the canonical maps
    of \eqref{diag:atomicity} -- with $X = \eta(x)$ and $A = \eta(a)$
    -- are isomorphisms.
  \item $\SP[\A]$ is \emph{soft}, meaning that the canonical diagrams
    of \eqref{diag:softness} are pushouts.
  \end{enumerate}      
  Moreover, if $\B$ is a $\SP$-category with a functor $F : \A \rTo
  \B$, so that the pair $(F,\B)$ satisfies 1,2,3, then the extension
  $\hat{F} : \SP[\A]\rTo \B$ is an equivalence of categories.
\end{theorem}
Thus, the structure of the category $\SP(\A)$ is precisely
determined by the conditions 1,2,3. We shall spend the next section
giving an explicit account of the property of 
softness. The theorem is a special instance of the more general
observations due to Joyal on free bicomplete categories \cite{joyal1,joyal2}.


\section{An account of softness}
\label{sec:softness}
A decision procedure necessarily focuses on the homset
$\Hom{X_{0}\times X_{1},A_{0} + A_{1}}$ which, by Theorem
\ref{theo:characterization}, is a certain the pushout.
Equivalently, this homset 
is the colimit of what we shall refer to as the ``diagram of
cardinals'':
$$
\xymatrix @C+2pt{%
  {\Hom{X_{0} \times X_{1},A_{0}}} & &  \Hom{X_{0},A_{0}} 
  \ar@/-2mm/[ll]_{\pi_0} 
  \ar@/-2mm/[rr]^{\sigma_{0}} & &  \Hom{X_{0},A_{0}+A_{1}}\\
  & & & & \\
  \Hom{X_{1},A_{0}}  \ar[uu]^{\pi_1} 
  \ar[dd]_{\sigma_{0}} & & & & 
  \Hom{X_{0},A_{1}}  \ar[dd]_{\pi_0} 
  \ar[uu]^{\sigma_{1}}\\
  & & & & \\
  \Hom{X_{1},A_{0}+A_{1}} & & \Hom{X_{1},A_{1}} \ar@/-2mm/[rr]^{\pi_1} 
  \ar@/-2mm/[ll]_{\sigma_{1}}
  & & \Hom{X_{0} \times X_{1},A_{1}}
}
$$
The explicit way of constructing such a colimit -- see \cite[\S
V.2.2]{maclane} -- is to first consider the sum $S$ of the corners:
$$
  \Hom{X_{0},A_{0}+A_{1}} + \Hom{X_{1},A_{0}+A_{1}} 
  + \Hom{X_{0}
    \times X_{1},A_{0}} + \Hom{X_{0} \times X_{1},A_{1}}
$$
and then quotient $S$ by the equivalence relation generated by
elementary pairs, i.e. pairs
 $(f,g)$ such that, for some $h$, $f =
\pi_{i}(h)$ and $\sigma_{j}(h) = g$, as sketched below:
$$
\xymatrix{%
  & h \in \Hom{X_{i}, A_{j}}
  \ar[ld]_{\pi_{i}} \ar[rd]^{\sigma_{j}}
  \\
  f \in \Hom{X_{0} \times X_{1}, A_{j}}
  &&
  g \in \Hom{X_{i}, A_{0} + A_{1}}
}
$$
Thus, for $f,f' \in S$ we have that $[f] = [f'] \in \Hom{X_{0}\times
  X_{1},A_{0} + A_{1}}$ if and only if there is a path in the diagram
of cardinals from $f$ to $f'$, that is a sequence 
$f_{0}f_{1}f_{2}\ldots f_{n}$, where $f = f_0$ and $f_n = f'$, such that, 
for $i = 0,\ldots ,n-1$, $(f_{i},f_{i+1})$ or $(f_{i+1},f_{i})$ is an 
elementary pair.


\section{The geometry of softness: weak disjointeness}
\label{sec:weakdisjointness}

Let us recall that a \emph{point} in a $\SP$-category is an arrow of
the form $p : 1 \rTo A$. When an object has a point we shall say it is 
\emph{pointed}.  Similarly, a \emph{copoint} is an arrow of
the form $c : X \rTo 0$ and an object with a copoint is 
\emph{copointed}.

An object of $\SP[\emptyset]$ can be viewed as a two-player game on a
finite tree, with no draw final position. Points then correspond then
to winning strategies for the player, while copoints correspond to
winning strategies for the opponent. Thus, by determinacy, every
object of $\SP[\emptyset]$ either has a point or a copoint but not
both.

The first important result for analyzing softness concerns
copoints and coproduct injections:

\begin{wrapfigure}[8]{r}{0pt}
  $$
    \xygraph{%
      []*+{X}
      (
      :[rrr]*+{A_{1}}="NE"^{g}
      :[dd]*+{A_{0} + A_{1}}="E"^{\sigma_{1}}
      )
      (:[dd]*+{A_{0}}="SO"_{f}
      :"E"^{\sigma_{0}}
      )
      :@{{}{.}{>}}[d(0.8)r(0.6)]*+{0}^{\exists  c}
      (:"NE"|{?_{A_{1}}},:"SO"|{?_{A_{0}}})
    }
    $$
\end{wrapfigure}
\myvspace{-12pt}\par
\begin{theorem}
  \label{theo:weakdisjointness}
   Coproducts are, in $\SP[\A]$, weakly disjoint: if $f;\sigma_{0} =
  g;\sigma_{1} : X \rTo A + B$, then there exists a copoint $c : X
  \rTo 0$ such that $f = c ; ?$ and $g = c; ?$.
\end{theorem}
The property is illustrated in the diagram.  The Theorem has
an interesting interpretation from the perspective of processes: a
process can send incoherent messages -- white noise -- on a channel 
without changing the meaning of the communication when and only when the  
recipient has stopped listening.  The consequences of misjudging 
when the recipient stops listening, of course, is well-understood by
school children and adults alike!

\begin{proof}
  We sketch here the proof of the
  Theorem  \ref{theo:weakdisjointness},
  emphasizing its geometrical flavor, as the diagram of cardinals is a
  sort of a one dimensional sphere.

  We say that a triple $(X\mid A_{0},A_{1})$ is good if for every $f:
  X \rTo A_{0}$ and $g : X \rTo A_{1}$ the statement of the Theorem
  holds. Similarly, we say that a triple $(X_{0},X_{1}\mid A)$ is good
  if, for every $f : X_{0} \rTo A$ and $g : X_{1} \rTo A$, the dual
  statement of the Theorem holds. We prove that every triple is good,
  by induction on the structural complexity of a triple.

  The non trivial induction step arises when considering a triple of
  the form $(X_{0} \times X_{1}\mid A_{0},A_{1})$ -- or the dual
  case. Here, saying that the equality $f;\sigma_{0} = g;\sigma_{1}$
  holds means that there exists a path $\phi$ of the form
  $f_{0}f_{1}\ldots f_{n}$ in the diagram of cardinals from $f=f_{0}
  \in \hom(X_{0}\times X_{1},A_{0})$ to $g = f_{n}\in \hom(X_{0}\times
  X_{1},A_{1})$, i.e. from northwest to southeast.  Moreover, we may
  assume $\phi$ to be simple.

  Such path necessarily crosses one of southwest or northeast
  corners, let us say the latter. This means that, for some $i =
  1,\ldots ,n-1$, $f_{i} \in \Hom{X_{0},A_{0} + A_{1}}$, and
  $f_{i-1},f_{i+1}$ are in opposite corners. W.l.o.g. we can assume
  $f_{i -1} \in \Hom{X_{0}\times X_{1},A_{0}}$ and $f_{i + 1} \in
  \Hom{X_{0}\times X_{1},A_{1}}$. Taking into account the definition
  of an elementary pair, we see that for some $h \in
  \Hom{X_{0},A_{0}}$ and $h' \in \Hom{X_{0},A_{1}}$ we have
  $h;\sigma_{0} = f_{i} = h'\sigma_{1}$. Thus, by the inductive
  hypothesis on $(X_{0}\mid A_{0},A_{1})$, we have $h = c;?_{A_{0}}$
  and $h' = c;?_{A_{1}}$; in particular the projection $\pi_{0}: X_{0}
  \times X_{1} \rTo X_{0}$ is epic, because of the existence of a
  copoint $c : X_{0} \rTo 0$. Recalling that the path $\phi$ is
  simple, we deduce that $i$ is the only time $\phi$ visits northeast,
  i.e. such that $f_{i} \in \Hom{X_{0},A_{0}+ A_{1}}$.


  A similar analysis shows that if $\phi$ crosses a corner, then it
  visits that corner just once. Thus, we deduce that $\phi$ does not
  cross the northwest corner, as $\phi$ visits the northwest corner 
  at time $0$ and a corner may be crossed only at time $i \in \set{1,\ldots
    ,n-1}$. Similarly, $\phi$ does not cross the southeast corner. Also, $\phi$
  cannot visit the southwest corner, as this would imply that at least one of
  northwest or southeast corners has been crossed.

  Putting these considerations together, we deduce that $\phi$ visits
  the northwest, northeast, and southeast corners exactly once. That
  is, $\phi$ has length $2$ and $i =1$.  Recalling the definition of
  elementary pair, we have
  \begin{align*}
    f = f_{0} & = \pi_{0};h \,, &
    h;\sigma_{0} & = f_{1}\,, &
    f_{1} & = h';\sigma_{1}\,, &
    \pi_{0};h' & = f_{2} = g\,.
  \end{align*}
  Considering that $h = c;?_{A_{0}}$ and $h' = c;?_{A_{1}}$, we deduce
  that $f =\pi_{0};c;?_{A_{0}}$ and $g =\pi_{0};c;?_{A_{1}}$.  \qed
\end{proof}

There are a number of consequences of this Theorem 
relevant to the decision procedure. To this end we need to introduce
some terminology and some observations.
We say that an arrow $f$ is \emph{pointed} if it factors through a
point, i.e. if $f = !;p$ for some point $p$. Similarly, an arrow is
\emph{copointed} if it factors through a copoint. Note that an object $A$ is
pointed iff $? : 0 \rTo A$ is pointed and, similarly, $X$ is copointed iff
$!: X \rTo 1$ is copointed.  A map which is neither pointed now copointed is 
said to be \emph{definite}, otherwise it is said to be \emph{indefinite}.

The following two facts are consequences of the theorem which can be 
obtained by a careful structural analysis:

\begin{corollary}~
  \label{cor:corollaries}
  \begin{enumerate}
  \item \label{item:copointed} It is possible to decide (and find
    witnesses) in linear time in the size of a term whether it
    is pointed or copointed.
  \item \label{item:monic} A coproduct injection $\sigma_{0} : A
    \rTo[\;\;] A + B $ is monic iff either $B$ is not pointed or $A$
    is pointed. In particular $? : 0 \rTo[\;\;] B$ is monic iff $B$
    is not pointed.
  \end{enumerate}
\end{corollary}

An arrow is a \emph{disconnect} if it is both pointed and copointed:
it is easy to see that there is at most one disconnect between any two
objects.  Furthermore, if an arrow $f: A \rTo B$ is copointed, that is
$f = c;?$, and its codomain, $B$, is pointed then $f$ is this unique
disconnect.  On the other hand, if the codomain $B$ is not pointed
then $? : 0 \rTo B$ is monic and, thus, such an $f$ corresponds
precisely to the copoint $c$.  These observations allow the equality
of indefinite maps, i.e. pointed and copointed, to be decided in
linear time. 

\smallskip

A further important fact which also follows from \ref{cor:corollaries}, 
in a similar vein to the above, concerns whether a map in $\SP[\A]$ factors 
through a projection or a coprojection.  This can also be decided in linear 
time on the size of the term. This is by a structural analysis which 
we now sketch.

Suppose that we wish to determine whether $f = \sigma_0(f'): A \rTo B+C$.
If syntactically $f$ is $\sigma_1(f')$ then, as a 
consequence of Theorem \ref{theo:weakdisjointness}, the only way it can 
factorize is if the map is copointed.  However, whether $f$ is copointed 
can be determined in linear time on the term by Corollary 
\ref{cor:corollaries}.  The two remaining possibilities are that 
$f$ is syntactically $\{ f_1,f_2\}$ or $\pi_i(f')$.  In the former case, 
inductively, both $f_1$ and $f_2$ have to factorize through $\sigma_0$. 
In the latter case, when the map is not copointed, $f'$ itself must factorize 
through $\sigma_0$.

There is, at this point, a slight algorithmic subtelty: to determine whether 
$f$ can be factorized through a projection it seems that we may have to 
repeatedly recalculate whether the term is pointed or copointed and this 
recalculation would, it seems, push us beyond linear time. However, it is 
not hard to see that this the recalculation can be avoided simply by 
processing the term initially to include this information into the structure 
of the term (minimally two extra bits are needed at each node to indicate 
pointedness and copointedness of the map): subsequently this information 
would be available at constant cost.  The cost of adding this information 
into the structure of the term is linear and, even better, the cost of 
maintaining this information, as the term is manipulated, is a constant 
overhead.


\section{Bouncing}
\label{sec:bouncing}

Given the previous discussion, equality for indefinite terms is
understood and so we can focus our attention on definite terms.  The main
difficulty of the decision procedure concerns equality in the
homset $\Hom{X_{0} \times X_{1}, A_{0} + A_{1}}$. However, 
the proof of Theorem \ref{theo:weakdisjointness} has revealed an 
important fact: \emph{if two
  terms in this homset have a definite denotation, then any path in
  the diagram of cardinals that witnesses the equality between them
  cannot cross a corner of the diagram}; that is, such a path must
\emph{bounce} backward and forward on one side:
\begin{align}
  \label{eq:spanofhoms}
  \Hom{X_{0} \times X_{1},A_{j}} \lTo[\;\;\pi_{i}\;\;]
  \Hom{X_{i},A_{j}} \rTo[\;\;\sigma_{j}\;\;] \Hom{X_{i},A_{0} +
    A_{1}}\,.
\end{align}
In other words, in order to understand definite maps we need to study
the pushouts of the above spans. Notice that the proof of Theorem
\ref{theo:weakdisjointness} also reveals that some simple paths in the
diagram of cardinals have bounded length. However, that proof does not
provide a bound for the length of paths that bounce on one
side. It is the purpose of this section to argue that such a bound
does indeed exist and to explore the algorithmic consequences.

We start our analysis by considering a general span $B \lTo[f] A
\rTo[g]C$ of sets and by recalling the construction of its colimit, the
pushout $B +_A C$.
This can be constructed by subdividing $B$ and $C$ into the image of
$A$ 
and the complement of that image.  Thus, if $B = {\sf Im}(f) + B'$ and
$C = {\sf Im}(g) + C'$ then $B +_A C = A'+ {\sf Im}(\rho) + B'$ where
$\rho: A \to B +_A C$.  The image ${\sf Im}(\rho)$ is the quotient of
$A$ with respect to the equivalence relation witnessed by ``bouncing
data''; \emph{bouncing data} is a sequence of elements of $A$,
$(a_0,a_1,...,a_n)$, with, for each $0\leq i<n$ either $f(a_i)=
f(a_{i+1})$ or $g(a_i)= g(a_{i+1})$.
Bouncing data, $(a_0,a_1,...,a_n)$, is said to be \emph{irredundant} if
adjacent pairs in the sequence are identified for different reasons.
Thus, in irredundant bouncing data if $f(a_i) = f(a_{i+1})$ then
$f(a_{i+1}) \not= f(a_{i+2})$ and similarly for $g$.  Redundant
bouncing data can always be improved to be irredundant by simply
eliding intermediate redundant steps.

For bouncing data of length 2, $(a_0,a_{1},a_2)$, we shall write $a_1:
a_0 \leadsto a_2$ to indicate $f(a_0)=f(a_{1})$ and $g(a_{1})=g(a_2)$,
and we shall call $a_1$ a \emph{bouncer} from $a_0$ to $a_2$. The
following is a general observation concerning pushouts of sets:
\begin{proposition}
  \label{prop:commutingequivalences}
  For any pushout of $B \lTo[f] A
  \rTo[g]C$ in sets the following are equivalent:
  \begin{enumerate}
  \item If $a_{0},a_{n}$ are related by some bouncing data, then they
    are related by bouncing data of length at most 2.
  \item The equivalence relations generated by $f$ and $g$ commute.
  \item The pushout diagram is a weak pullback, i.e. the comparison
    map to the pullback is surjective.
  \end{enumerate}
  Moreover, when one of these equivalent conditions holds, the pushout
  is a pullback iff for every $a_0$ and $a_2$ related by bouncing data
  there is a {\em unique\/} element $a_1$ such that $a_1: a_0 \leadsto
  a_2$.
\end{proposition}

\vskip -24pt
\begin{wrapfigure}[10]{r}{0pt}
  $$\xymatrix{%
    & X_{i} \ar[r]^f \ar@{..>}[ddr]^{\exists ! \;h} & A_{j}  \ar[rd]^{\sigma_j} \\
    X_{0} \times X_{1} \ar[ru]^{\pi_i} \ar[rd]_{\pi_i} & & & A_{0}+A_{1} \\
    & X_{i} \ar[r]_g & A_{j} \ar[ru]_{\sigma_j} }$$ 
\end{wrapfigure}
\myvspace{12pt}\par
Surprisingly, this altogether special situation holds in $\SP(\A)$.
More precisely, we say that the homset $\Hom{X_{i},A_{j}}$ \emph{bounces} if,
for each pair of objects $X_{1-i},A_{1-j}$, the span \eqref{eq:spanofhoms}
has a pushout which makes the homset $\Hom{X_{i},A_{j}}$ the pullback.
Intuitively, $\Hom{X_{i},A_{j}}$ bounces if, whenever the upper and
lower legs of the diagram on the right are equal (and definite), this
is because of a unique bouncer $h:f \leadsto g$, where $h$ is shown
dotted and the fact that it is a bouncer means that the two smaller
rectangles commute.  Thus we have:
\begin{theorem}
\label{thm-bounce}
In $\SP(\A)$ all homsets bounce.
\end{theorem}
The Theorem implies that if $f$ and $g$ are related by a
bouncing  path in the diagram of cardinals, then there exists a path of
length at most 2 relating them.

The proof of the Theorem \ref{thm-bounce} relies on a tricky structural induction on
the pairs $(X_{i},A_{j})$. Rather than presenting it here, we shall illustrate
the proof for the special case of $\SP(\emptyset)$, the initial
$\SP$-category.  Here the situation is much simpler, as noted above, since each 
object is either pointed or copointed, but not both.
We observe first that when there is a map from $X_{i}$ to $A_{j}$, if
$X_{i}$ is pointed then $A_{j}$ must be pointed as well and, dually,
when $A_{j}$ is copointed $X_{i}$ must be copointed. As $X_{i}$ and
$A_{j}$ must be either pointed or copointed it follows that $X_{i}$ is
pointed (respectively copointed) if and only if $A_{j}$ is.  However,
if $A_{j}$ is pointed then $\sigma_j$ is monic so the bouncer $h$ is
forced to be $f$.  Otherwise, if $A_{j}$ is not pointed, then $A_{j}$
is copointed and $X_{i}$ as well; then $\pi_i$ is epic and the bouncer
$h$ is forced to be $g$.

When $h \in \set{f,g}$, say that the bouncers $h : f \leadsto g$ is trivial.  
While $\SP(\emptyset)$ has only trivial bouncers, the
next example shows that this not in general the case. Let $k : x \to
a$ be an arbitrary map of $\A$, let $X_{0} = (0 \times 0)+\eta(x)$ and $A_{0}
=(1+1) \times \eta(a)$, let $z : 0 \times 0 \rTo 1 \times 1$ be the
unique disconnect. Recalling that an arrow from a coproduct to a
product might be represented as a matrix, define
\begin{align*}
  f & = 
  \begin{mymatrix}
    z & \pi_{0}(\ecotuple{}) \\
    \sigma_{0}(\etuple{}) & \eta(k)
  \end{mymatrix}
  & 
  h & = \begin{mymatrix}
    z &  \pi_{1}(\ecotuple{}) \\
    \sigma_{0}(\etuple{}) & \eta(k)
  \end{mymatrix} 
  & 
  g & = \begin{mymatrix}
    z & \pi_{1}(\ecotuple{}) \\
    \sigma_{1}(\etuple{}) & \eta(k)
  \end{mymatrix}
\end{align*}
as arrows of the homset $\Hom{X_{0},A_{0}}$.  Then $h: f \leadsto g$
is an example of a non-trivial bouncer whenever $X_{1}$ is copointed
and $A_{1}$ is pointed, since then $f$ and $h$ are coequalized by
$\sigma_{0}$ and $h$ and $g$ are equalized by $\pi_0$.  Notice,
however, that this example relies crucially on having atomic
objects. Also, this is a sort of minimal example of a non trivial
bouncer; it 
suggested to us that the equivalence relations generated by
$\sigma_{0}$ and $\pi_{0}$ might commute, see Proposition
\ref{prop:commutingequivalences}.

\begin{wrapfigure}[12]{r}{0pt}
\begin{lstlisting}[mathescape]
let equivalent $f\;g$ = 
  let $f',g'$ be such that
    $f \semeq \sigma_{j}(f')$ and $g \semeq \pi_{i}(g')$
  in if $f'$,$g'$ do not exist then 
         false
     elseif $X_{1-i}$ is copointed then 
	 equal $\sigma_{j}(f')\;\sigma_{j}(g')$
     else (*$A_{1-j}$ is pointed*) 
         equal $\pi_{i}(f')\;\pi_{i}(g')$
\end{lstlisting}
\end{wrapfigure}
\myvspace{-12pt}\par
We conclude this Section by sketching an algorithm --- named
\cd{equivalent}, which we present on the right for
$\SP[\emptyset]$ -- that computes whether a term $f$ of the homset
$\hom(X_{0}\times X_{1},A_{j})$ is equivalent to a term $g$ of 
the homset $\hom(X_{i},A_{0}+A_{1})$ within the pushout of the span
\eqref{eq:spanofhoms}. The algorithm tries to lift $f$ and $g$ to
$f',g'$ in the homset $\hom(X_{i},A_{j})$ and, if successful, it tests
for the existence of a bouncer $h:f'\leadsto g'$. Notice that the
algorithm is defined by mutual recursion on the general decision 
procedure \cd{equal}.


\section{The decision procedure}
\label{sec:dp}

We present in Figure \ref{fig:dp} the decision procedure for
$\SP(\emptyset)$. The general decision procedure for $\SP(\A)$ --
which depends on having a decision procedure for $\A$ -- is
considerably complicated by having to construct non-trivial bouncers;
we describe it in the full paper.
\begin{figure}[t]
  \centering
\begin{lstlisting}[mathescape]
let equal $f\; g$ = match (dom $f$,cod $g$) with
    ($0$,_) | (_,$1$) -> true
  | ($1$,$A_0 + A_1$) -> 
      let $i,f',j,g'b$ be such that
	$f \semeq \sigma_i(f')$ and $g \semeq \sigma_j(g')$
      in if $i = j$ then equal $f' \; g'$ else false
  | ($Y_0 \times Y_1$,$0$) ->  ... dual
  | (_,$A_0 \times A_1$) -> 
      let $f_0,f_1,g_0,g_1$ be such that
	$f \semeq \langle f_0,f_1\rangle$ and $g \semeq \langle g_0,g_1\rangle$
      in (equal $f_0 \;g_0$) && (equal $f_1 \;g_1$)
  | ($Y_0 + Y_1$,_) -> ... dual
  | ($X_0 \times X_{1},A_0 + A_{1}$) -> 
      if definite $f\;g$ then
	    match ($f,g$) with 
		($\pi_{i}(f'),\sigma_{j}(g')$) | ($\sigma_{j}(g'),\pi_{i}(f')$) -> equivalent $f' \;g'$
	      | ($\pi_{i}(f'),\pi_{i}(g')$) -> 
		  let $i,\tilde{g}$ be such that
                    $\pi_{i}(g') \semeq \sigma_{j}(\tilde{g})$
                  in
                    if such $i,\tilde{g}$ do not exist then equal $f' \;g'$
                    else equivalent $f'\;\tilde{g}$
	      | ($\sigma_{i}(f'),\sigma_{i}(g')$) -> ... dual
	      |  _   -> false
      else equal_indefinite $f\;g$

\end{lstlisting}
  \caption{The decision procedure for $\SP[\emptyset]$.}
  \label{fig:dp}
\end{figure}

\smallskip

\noindent
\textbf{The procedure.}~
The procedure starts with two parallel terms in $\SP(\emptyset)$,
$f,g: X \rightarrow A$.  If $X$ is initial or $A$ is final then we are done
-- there are of course no maps if $X$ is final and $A$ is initial.  If
either $X$ is a coproduct or $A$ is a product we can decompose the
maps and recursively check the equality of the components.  Thus, if
$X = X_1 + X_2$ then $f = \{ \sigma_0;f,\sigma_1;f \}$ and $g = \{
\sigma_0;g,\sigma_1;g \}$, and then $f=g$ if and only if $\sigma_i;f =
\sigma_i;g$ for $i=0,1$.  This requires that one cut-eliminates the
compositions with $\sigma_i$ -- which can be performed in time linear
in the size of the term.

\smallskip
This reduces the problem to the situation in which the domain of the
maps is a product and the codomain is a coproduct.  Here we have to
consider two cases:

\smallskip

\emph{Indefinite maps.}~
In section \ref{sec:weakdisjointness} we mentioned that in time linear
on the size of the maps (which is in turn bounded by the product of
the types) one can determine whether the map is pointed (and produce a
point) or copointed (and produce a copoint).  If both terms are
pointed and copointed then they are the unique disconnect and we are
done.  If one term is just pointed the other must be just pointed and
the points must agree (and dually for being just copointed).

\smallskip

\emph{Definite maps.}~%
When the maps are definite then a first goal is to determine whether
the term $f$ factors through a projection or a coprojection or,
indeed, both (i.e. $f = \sigma_i(f')$ or $f = \pi_j(f')$).  These
factoring properties, as was discussed above, can be determined in
linear time.  Using these 
properties -- remembering that a path in the diagram of cardinals that
relates two definite terms can only move along a side -- there are two
cases, either they bounces or they do not.  It they bounce we can
reduce the problem to the case when one term factors through a
projection and the other through a coprojection (using
\cd{equivalent}).  If the terms do not bounce then they both must
factor {\em syntactically\/} in the same manner so that $f$ is
$\sigma_0(f')$ and $g$ is $\sigma_0(g')$, then $f'$ must equal
$g'$. 

\smallskip

\noindent
\textbf{Complexity.}~
To obtain the complexity of this algorithm we shall use  an 
important observation: %
\emph{%
  in $\SP[\emptyset]$ the size of any cut-eliminated term 
  representing an arrow $t: X \rTo  A$ is bounded by the 
  product of the sizes of the types and its height is bounded by the 
  sum of the heights of the types. %
}%
This is proven by a simple structural induction.

The decision procedure now uses one preprocessing sweep to annotate the terms 
(and the types) with information concerning what is pointed and copointed.  
Then the main equality algorithm is applied which employs two sorts of 
algorithm (on subterms), which manipulate the terms and require 
linear time on the maximal size of the input and output terms.  

The first of these algorithm simply forms a tuple when the codomain is
a product and a cotuple when the domain is a sum.  The second
algorithm determines whether a term can be factored via a projection
or coprojection and returns a factored version.  Getting this to run
in linear time does require that the pointed and copointed information
can be retrieved in constant time (which is managed by preprocessing
the terms).

The other major step in the algorithm, which we have not discussed for the 
general case, involves finding a bouncer.  In the $\SP(\emptyset)$ case 
this involves determining which of the projection or coprojection is 
respectively epic or monic.  This, in turn, is determined by the pointedness 
or copointedness of the components of the type which can usefully be 
calculated in the preprocessing stage -- and so is constant time.

Essentially this means that the algorithm at each node of the term 
requires processing time bounded by a time proportional to the (maximal) size of 
the subterm.  Such a pattern of processing is bounded by time proportional 
to the height of the term times the size.  We therefore have:
\begin{proposition}
  To decide the equality of two parallel terms $t_1,t_2: A \rTo B$ in
  $\SP(\emptyset)$ has complexity in ${\cal O}(({\sf hgt}(A)+{\sf
    hgt}(B)) \cdot {\sf size}(A) \cdot {\sf size}(B))$.
\end{proposition}
The analysis of the algorithm for $\SP(\A)$ is more complex and is left 
to the fuller exposition.

\bibliographystyle{splncs}
\bibliography{../biblio}

\begin{thebibliography}{10}

\bibitem{lambek}
Lambek, J.:
\newblock Deductive systems and categories. {I}. {S}yntactic calculus and
  residuated categories.
\newblock Math. Systems Theory \textbf{2} (1968)  287--318

\bibitem{dosen_book}
Do{\v{s}}en, K.:
\newblock Cut elimination in categories. Volume~6 of Trends in Logic---Studia
  Logica Library.
\newblock Kluwer Academic Publishers, Dordrecht (1999)

\bibitem{CS00}
Cockett, J.R.B., Seely, R.A.G.:
\newblock Finite sum-product logic.
\newblock Theory Appl. Categ. \textbf{8} (2001)  63--99 (electronic)

\bibitem{joyal_hu}
Hu, H., Joyal, A.:
\newblock Coherence completions of categories.
\newblock Theoret. Comput. Sci. \textbf{227}(1-2) (1999)  153--184 Linear
  logic, I (Tokyo, 1996).

\bibitem{glabbeek}
Hughes, D.J.D., van Glabbeek, R.J.:
\newblock Proof nets for unit-free multiplicative-additive linear logic
  (extended abstract).
\newblock In: LICS, IEEE Computer Society (2003)  1--10

\bibitem{dosen_last}
Do{\v{s}}en, K., Petri{\'c}, Z.:
\newblock Bicartesian coherence revisited.
\newblock In Ognjanovic, Z., ed.: Logic in Computer Science, Zbornik Radova.
  Volume~12.
\newblock (2009) arXiv:0711.4961.

\bibitem{blass}
Blass, A.:
\newblock A game semantics for linear logic.
\newblock Ann. Pure Appl. Logic \textbf{56}(1-3) (1992)  183--220

\bibitem{joyal_money_games}
Joyal, A.:
\newblock Free lattices, communication and money games.
\newblock In: Logic and scientific methods (Florence, 1995). Volume 259 of
  Synthese Lib.
\newblock Kluwer Acad. Publ., Dordrecht (1997)  29--68

\bibitem{luigis}
Santocanale, L.:
\newblock Free {$\mu$}-lattices.
\newblock J. Pure Appl. Algebra \textbf{168}(2-3) (2002)  227--264 Category
  theory 1999 (Coimbra).

\bibitem{pastro2}
Cockett, J.R.B., Pastro, C.A.:
\newblock A language for multiplicative-additive linear logic.
\newblock Electr. Notes Theor. Comput. Sci. \textbf{122} (2005)  23--65

\bibitem{pastro3}
Cockett, J.R.B., Pastro, C.A.:
\newblock The logic of message passing.
\newblock CoRR \textbf{abs/math/0703713} (2007)

\bibitem{maclane}
Mac~Lane, S.:
\newblock Categories for the working mathematician. Second edn. Volume~5 of
  Graduate Texts in Mathematics.
\newblock Springer-Verlag, New York (1998)

\bibitem{joyal1}
Joyal, A.:
\newblock Free bicomplete categories.
\newblock C. R. Math. Rep. Acad. Sci. Canada \textbf{17}(5) (1995)  219--224

\bibitem{joyal2}
Joyal, A.:
\newblock Free bicompletion of enriched categories.
\newblock C. R. Math. Rep. Acad. Sci. Canada \textbf{17}(5) (1995)  213--218

\end{thebibliography}

\end{document}